\documentclass[prb,twocolumn]{revtex4}
\usepackage[utf8]{inputenc}

\usepackage{graphicx}
\usepackage{subfigure}
\usepackage{textcomp}
\usepackage{amsmath}
\usepackage{enumitem}
\usepackage{sidecap}

\newcommand{\nvm}{NV$^{-}$}
\newcommand{\nvz}{NV$^{0}$}
\newcommand{\tat} {$^2A_2$}
\newcommand{\te}{$^2E$}

\newcommand{\ABS}{Z_{L+R}}
\newcommand{\MCD}{Z_{L-R}}
\newcommand{\bra}[1] {\langle #1 |}
\newcommand{\ket}[1] {| #1 \rangle}

\newcommand{\me}[3] {\bra{#1} #2 \ket{#3}}

\newcommand{\ms}[1]{\left< #1 \right>}
\newcommand{\lparam}{0.0186}
\newcommand{\lparamse}{0.0005}
\newcommand{\soparam}{2.24}
\newcommand{\soparamse}{0.05}

\begin{document}

\title{The fine structure of the neutral nitrogen-vacancy center in diamond}

\author{M.S.J. Barson$^{1}$}
\author{E. Krausz$^{2}$}
\author{N.B. Manson$^1$}
\author{M.W. Doherty$^1$}

\affiliation{$^1$Laser Physics Centre, Research School of Physics and Engineering, Australian National University, Acton, 2601, Australia}
\affiliation{$^2$Research School of Chemistry, Australian National University, Acton, 2601, Australia}
\altaffiliation{marcus.doherty@anu.edu.au}

\begin{abstract}
	The nitrogen-vacancy (NV) center in diamond is a widely-utilized system due to its useful quantum properties. Almost all research focuses on the negative charge state (\nvm) and comparatively little is understood about the neutral charge state (\nvz). This is surprising as the charge state often fluctuates between \nvz\, and \nvm\, during measurements. There are potentially under utilized technical applications that could take advantage of \nvz, either by improving the performance of \nvm\, or utilizing \nvz\, directly. However, the fine-structure of \nvz\, has not been observed. Here, we rectify this lack of knowledge by performing magnetic circular dichroism (MCD) measurements that quantitatively determine the fine-structure of \nvz. The observed behavior is accurately described by spin-Hamiltonians in the ground and excited states with the ground state yielding a spin-orbit coupling of $\lambda = \soparam \pm \soparamse$ GHz and a orbital $g-$factor of $\lparam \pm \lparamse$. The reasons why this fine-structure has not been previously measured are discussed and strain-broadening is concluded to be the likely reason.
\end{abstract}

\maketitle

\section{Introduction}

The nitrogen-vacancy (NV) center is a promising color center in diamond for quantum technology. Applications include nanoscale quantum sensing and quantum information processing. This remarkable utility is due to the useful spin dependent photodynamics of the center's negative charge state (\nvm). Despite the intense interest in \nvm, surprisingly little is known about the other common charge state of the NV center, the neutrally-charged \nvz. It is important to understand \nvz\, for a number of reasons. In particular, there is a long standing puzzle as to why the ground state electron paramagnetic resonance (EPR) signal \cite{PhysRevB.77.081201} has not been seen in either \nvz\, or similar $^2E$ ground state centers, such as the negatively-charged silicon-vacancy \cite{SiV0_2011_EPR,kubanek-haussler-photo-siv-gev-2017,1367-2630-17-4-043011} (SiV$^-$) and the negatively-charged germanium-vacancy \cite{kubanek-haussler-photo-siv-gev-2017} (GeV$^-$) centers. 

There are similarities \cite{Davies_1979,rogers2015singlet} between the Jahn-Teller (JT) induced vibronic structure of the \nvz\, ground state and of the poorly understood singlet states of \nvm. The properties of these singlet states and their associated inter-system crossings (ISC) underpin all of the desirable spin-dependent photo-dynamics of \nvm, such as spin-polarization and readout. Greater understanding of the \nvm\, singlet states will lead to the discovery of ways to enhance these properties of \nvm\, or the ability engineer other defects with improved properties. Due to their similarities, the clues to understanding the behavior of the \nvm\, singlet levels may lead from a better understanding of the \nvz\, ground state.

There are also technical applications utilizing \nvz\, in conjuncture with \nvm. One application is controlling the NV charge state to limit the dephasing effect of the electron spin on nearby nuclear spins \cite{pfender2017nv+}. A better understanding of the ground state electron spin of \nvz\, could also lead to improved control and longer nuclear spin coherence. Another application is the ability to detect the electron ejected to the conduction band during ionization from \nvm\, to \nvz\, for spin-to-charge readout of the \nvm\, spin \cite{Bourgeois2015}. This is used as an alternative spin readout process for quantum applications utilizing \nvm. A further application is using the long-lived photoionization of \nvm\, to \nvz\, for classical data storage \cite{Dhomkare1600911}.

\begin{figure}
	\centering
	\includegraphics[width=0.4\textwidth]{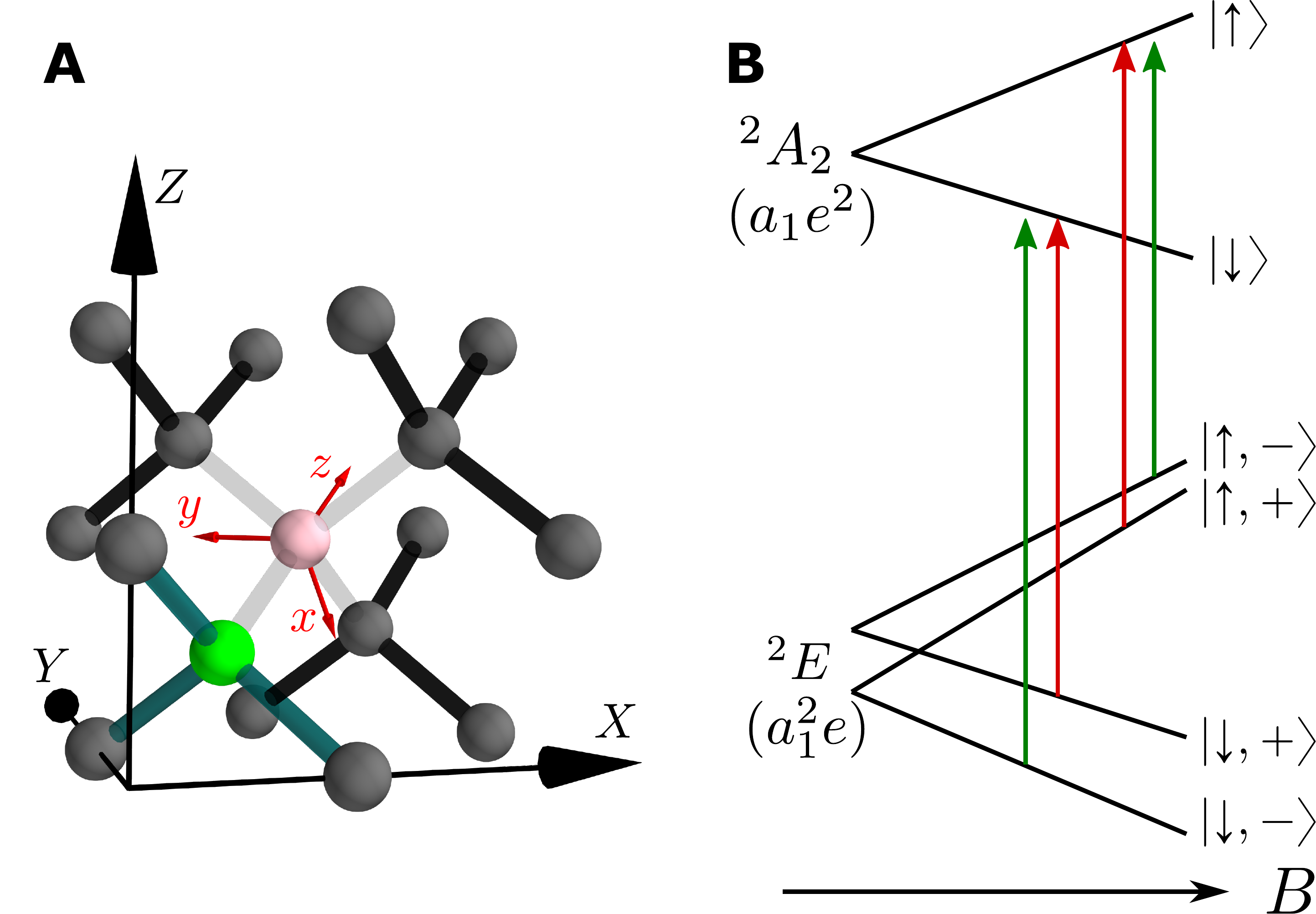}
	\caption{\textbf{(A)} Unit cell of diamond containing an NV center with crystallographic coordinates $(X,Y,Z) = ([100],[010],[100])$ and NV coordinates chosen as $(x,y,z)=([11\bar{2}],[\bar{1}10],[111])$. The carbon atoms are depicted as gray, the nitrogen atom as green and the vacancy site as pink. \textbf{(B)} Energy levels of the \nvz\, ground \te\,($a_1e^2$) and excited \tat\,($a_1^2e$) states, where the brackets signify the molecular orbitals \cite{doherty2011negatively}. The orbital states are denoted by  $+,\,-$ and spin states by $\uparrow,\,\downarrow$. Left and right circularly polarized optical transitions are denoted by the red and green arrows. The gradients of the ground state resonances with respect to magnetic field are unequal due to the orbital Zeeman competing with spin Zeeman effects.}
	\label{fig1}
\end{figure}

The only previously detected EPR \cite{PhysRevB.77.081201} of \nvz\, was for the metastable quartet state $^4A_2$. Since no ground or excited fine structure has been determined for \nvz\, either by photo-luminescence excitation (PLE) or by EPR, an alternative means of measurement is required. The optical resonance of \nvz\, is well known \cite{Davies_1979}. This provides a starting point and by probing the magneto-optical properties of this resonance, new information regarding the electronic structure of both ground and excited states can be obtained. One such magneto-optical measurement is magnetic circular dichroism (MCD) spectroscopy. MCD is a differential optical spectroscopic technique that determines the difference of left and right circularly polarized light in the presence of an axial magnetic field \cite{piepho1983mdc,Stephens_1974}. This provides quantitative measurements of the magnetic behavior of optical transitions. In particular, the fine structure of the ground and excited states can be extracted from the MCD measurements.

In this paper, we determine the fine structure of \nvz\, and develop a simple electronic model that fully describes the observed behavior. The MCD experiment is introduced and it is shown how the electronic model relates the resulting spectra. The results and model allow for the fine structure parameters of the system to be determined. We then discuss why such fine-structure behavior has not been observed in previous measurements.
\\

\section{Theory}
The NV center consists of a substitutional nitrogen impurity and an adjacent vacancy. The defect in its neutral charge state has 5 electrons, 3 from the dangling carbon bonds and 2 from the dangling nitrogen bond. The negatively charged state gains an additional electron from nearby charge donors (usually nitrogen) \cite{neil2018nv-n-pair-1b,manson2005photo}. The NV axis (line between nitrogen and vacancy sites) is aligned along a $\langle 111 \rangle$ direction and the defect has three-fold rotational symmetry ($C_{3v}$), as shown in figure \ref{fig1}(A).

The electronic structure of \nvz\, consists of a $^2E$ ground state, $^2A_2$ optical excited level and an intermediate meta-stable $^4A_2$ level. The zero phonon line (ZPL) of the $^2E\leftrightarrow$$^2A_2$ occurs at 2.16 eV (575 nm). The $^2A_2$ has no orbital degeneracy and exhibits no zero field fine structure. The ground state $^2E$ has orbital degeneracy that gives rise to spin-orbit fine structure as shown in figure \ref{fig1}(B). The \te\, fine structure can be described by the Hamiltonian
\begin{align}
H = g \mu_B \vec{S}\cdot\vec{B} + l \mu_B {L_z} B_z + 2\lambda {L_z} {S_z}, \label{eqn-Hamiltian}
\end{align}
where $\mu_B$ is the Bohr magneton, $g\sim2$ is the spin $g$-factor and $l$ is the orbital $g$-factor, $\lambda$ is the spin-orbit interaction parameter, the orbital $L_i = \sigma_i$ and spin $S_i = \frac{1}{2}\sigma_i$ operators are defined respectively in terms of the Pauli matrices $\sigma_i$ and $z$ is along the NV axis (see definition of coordinate system in figure \ref{fig1}(A)).

In the presence of a large magnetic field, it is convenient to transform the spin coordinate system so that $S_z$ is parallel to the applied field (i.e. $S_z \rightarrow S_z \cos \theta + S_x \sin \theta$), whilst the orbital operators remain unchanged because the orbitals remain defined by the crystal axes. The transformed Hamiltonian is then
\begin{align}
\begin{split}
H = \mu_B g {S_z} B + \mu_B l {L_z} B \cos\theta \\
+ 2\lambda {L_z} \left( {S_z} \cos \theta + {S_x} \sin \theta \right),
\label{eqn-Hamiltonian-secular}
\end{split}
\end{align}
For our measurements, the diamond sample had a $\ms{100}$ face that was aligned parallel to the magnetic field, and thus $\cos\theta =\hat{z}\cdot\hat{Z}= 1/\sqrt{3}$. For this field orientation, all possible NV orientations have an equivalent magnetic field projection. The eigenenergies of the transformed Hamiltonian are thus
\begin{align}
	E_{\pm\downarrow} &= \pm \mu_B l B \cos \theta - \tfrac{1}{2}\sqrt{4\lambda^2+(g\mu_B B)^2 \pm 4 \lambda g \mu_B B \cos\theta}  \nonumber \\
	E_{\pm\uparrow} &= \pm \mu_B l B \cos \theta + \tfrac{1}{2}\sqrt{4\lambda^2+(g\mu_B B)^2 \pm 4 \lambda g \mu_B B \cos\theta,}
\end{align}
where the $\uparrow,\downarrow$ signifies spin-up.-down states and the $\pm$ signifies orbital states with well-defined positive and negative units of orbital angular momentum about the NV axis. The optical selection rules are such that the $\pm$ orbital states participate in $R/L$ circularly polarized transitions respectively. Care was taken to transform the circularly polarized light from the lab coordinate system into the NV coordinate system, see supplementary information.

\begin{SCfigure*}
	\centering
	\includegraphics[width=0.6\textwidth]{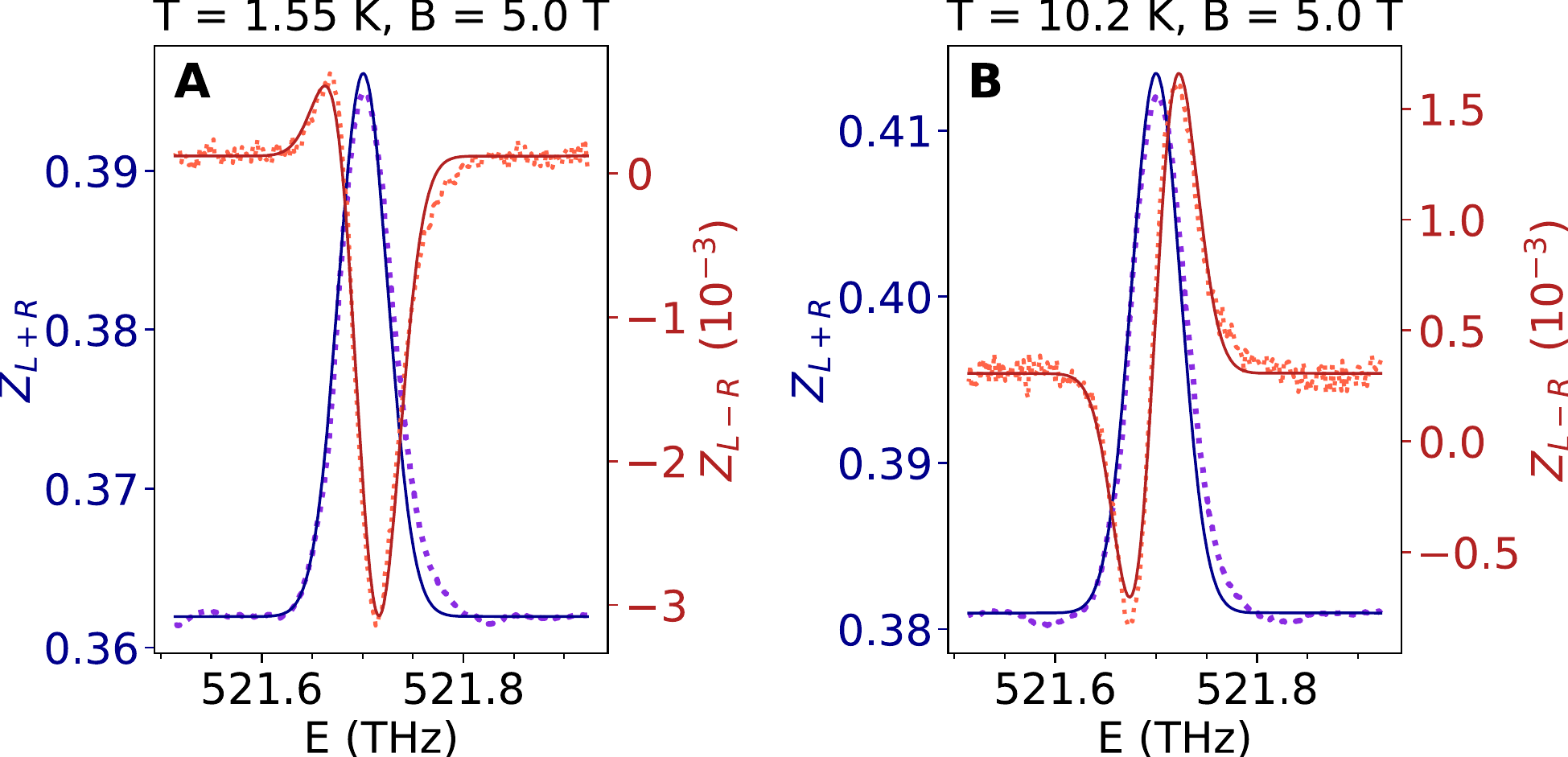}
	\caption{Examples of total ($\ABS$) and differential ($\MCD$) absorption spectra at B = 5 T and T = 1.55 K \textbf{(A)} and at B = 5 T and T = 10.2 K \textbf{(B)}. Notice the change in sign of the $\MCD$ signal for the higher temperature data. Plots are in units of absorbance $A = \log_{10}\frac{\Phi_i}{\Phi_t}$, where $\Phi_{i,t}$ are the incident and transmitted radiant fluxes. The fits are of the same form as shown in equation (\ref{eqn-Gaussian-models}).}
	\label{fig2}
\end{SCfigure*}

We use MCD to determine the magnetic splitting of the \te\, state. This allows for the spin-orbit parameter $\lambda$ and orbital $g-$factor $l$ to be determined. Since these measurements are performed at low temperature and high magnetic fields $(k_B T \sim g \mu_B B)$, the thermal populations of the \te\, ground state sub-levels mus also be included into the model. The optical transition being measured is spin allowed and the energy levels of the excited \tat\, must also be considered. We have assumed a simple spin$-1/2$ system with a $g-$factor of 2.  Thus the excited state eigenenergies are simply $E_{A\uparrow\downarrow} = E_\mathrm{ZPL}\pm g \mu_B B/2$, where $E_\mathrm{ZPL}$ is the optical ZPL energy.

\section{Materials and Method}

The experimental apparatus is described in Ref. \cite{Elmars2013OpticalSpectPhoto}. White light is passed through a double monochromator so as to provide monochromatic light that can be swept in frequency. This light is then passed through a chopper wheel to modulate the intensity (488 Hz modulation frequency). Next the light is filtered to remove second order diffracted light and is then linearly polarized and passed through a photoelastic modulator (PEM) which is driven at 50 kHz to create alternatively left and right circularly polarized light. The combination of the chopper wheel and periodic circularly polarized light allows for a heterodyne based detection process so as to measure small differences in left and right circularly polarized absorption. In this work, the differential absorption is about $10^2$ times weaker than the total absorption. The sample is placed in a temperature controlled (1.46 - 300 K) liquid helium immersion cryostat with a 0 - 6 T superconducting magnet.

The sample was a CVD grown (Element six) crystal with $\sim$ ppm concentration of nitrogen that was irradiated with electrons at a fluence of $10^{17}$/cm$^2$ and annealed at 750$^\text{o}$ for two hours to create  NV centers. The \nvz\, ZPL is on top of the \nvm\, absorption phonon sideband, however only the difference in absorption ZPL relative to the background is considered. As such, the results are not effect by the presence of the \nvm\, ZPL. The absorption of \nvz\, is weak with about a 7\% reduction in transmittance at the peak of the ZPL.

\section{Results}

The total absorption ($\ABS$) and the differential absorption ($\MCD$), as shown in figure \ref{fig2} are simultaneously acquired. The differential signal $\MCD=Z_L-Z_R$, is the difference of two non-degenerate absorptive lineshapes and the total absorption $\ABS=1/2(Z_L+Z_R)$ is the sum of the two absorptive lineshapes. By comparing the two spectra using either parametric curve fitting or moment analysis \cite{Stephens_1974,Lax1952momentsFrankCondon} the separation and magnitude of the two circularly polarized absorptions can be determined. Moment or curve fitting analyses each have separate benefits and drawbacks and both methods of analysis were pursued in parallel. For curve-fitting, Gaussian lineshapes were used with the free parameters $A_1, A_2, \sigma, \mu$ and $d$. $A_1$ and $A_2$ are the amplitudes of the lines, $2d$ is the separation of the two lines and $\mu$ and $\sigma$ are the shared central position and width of the two lines.
\begin{align}
Z_{L\pm R} (E) = \frac{A_1}{\sqrt{2\pi}\sigma}e^{\frac{(E-\mu-d)^2}{2\sigma^2}}\pm \frac{A_2}{\sqrt{2\pi}\sigma}e^{\frac{(E-\mu+d)^2}{2\sigma^2}}
\label{eqn-Gaussian-models}
\end{align}
The moment analysis is performed as described in Ref. \cite{piepho1983mdc}.

The first MCD quantity we consider describes the probability of absorbing a left or right circularly polarized photon. This can be expressed by the ratio of the zeroth $\ABS$ and $\MCD$ spectral moments or the amplitudes ($A_1$ and $A_2$) obtained from the parametric curve fitting
\begin{align}
	\frac{\langle Z_{L-R} \rangle_0}{\langle Z_{L+R} \rangle_0} &= 2\frac{\left(P_{-\downarrow} + P_{-\uparrow} \right) - \left(P_{+\downarrow} + P_{+\uparrow} \right)}{P_{-\downarrow} + P_{-\uparrow}+P_{+\downarrow} + P_{+\uparrow}}  \nonumber \\
	& = \frac{A_1-A_2}{A_1+A_2}. \label{eqn-Aratio}
\end{align}
In the above, the angle brackets $\langle Z_{L\pm R} \rangle_n$ represent the $n^{th}$ spectral moment. The second MCD quantity is probability weighted transition energies of absorbing a left or right circularly polarized photon. As such, it also provides information regarding the energy separation of possible the left and right circularly polarized transitions. This quantity can be determined from the ratio of the first and zeroth moments of the $\MCD$ and $\ABS$ spectra respectively. This quantity can also be determined by curve fitting using the energy separation ($2d$) and amplitudes ($A_1$ and $A_2$) of two lineshapes.
\begin{align}
		\frac{\langle Z_{L-R}\rangle_1}{\langle Z_{L+R}\rangle_0} &= \left[\left(P_{-\downarrow} (E_{-\downarrow}-E_{A\downarrow}) + P_{-\uparrow}(E_{-\uparrow}-E_{A\uparrow}) \right)\right. \nonumber \\
		&-\left.\left(P_{+\downarrow} (E_{+\downarrow}-E_{A\downarrow}) + P_{+\uparrow}(E_{+\uparrow}-E_{A\uparrow}) \right)\right] \nonumber \\
		&\times \frac{2}{P_{-\downarrow} + P_{-\uparrow}+P_{+\downarrow} + P_{+\uparrow}} \nonumber \\
	& = \frac{4dA_1 A_2}{(A_1+A_2)^2}. \label{eqn-d}
\end{align}
In the above, the probabilities $P_i$ are determined from the thermal occupations $P_i = e^{-E_i/k_B T}/\sum_j e^{-E_j/k_B T}$ of the fine structure states of the $^2E$ level.

\begin{SCfigure*}
	\centering
	\includegraphics[width=0.6\textwidth]{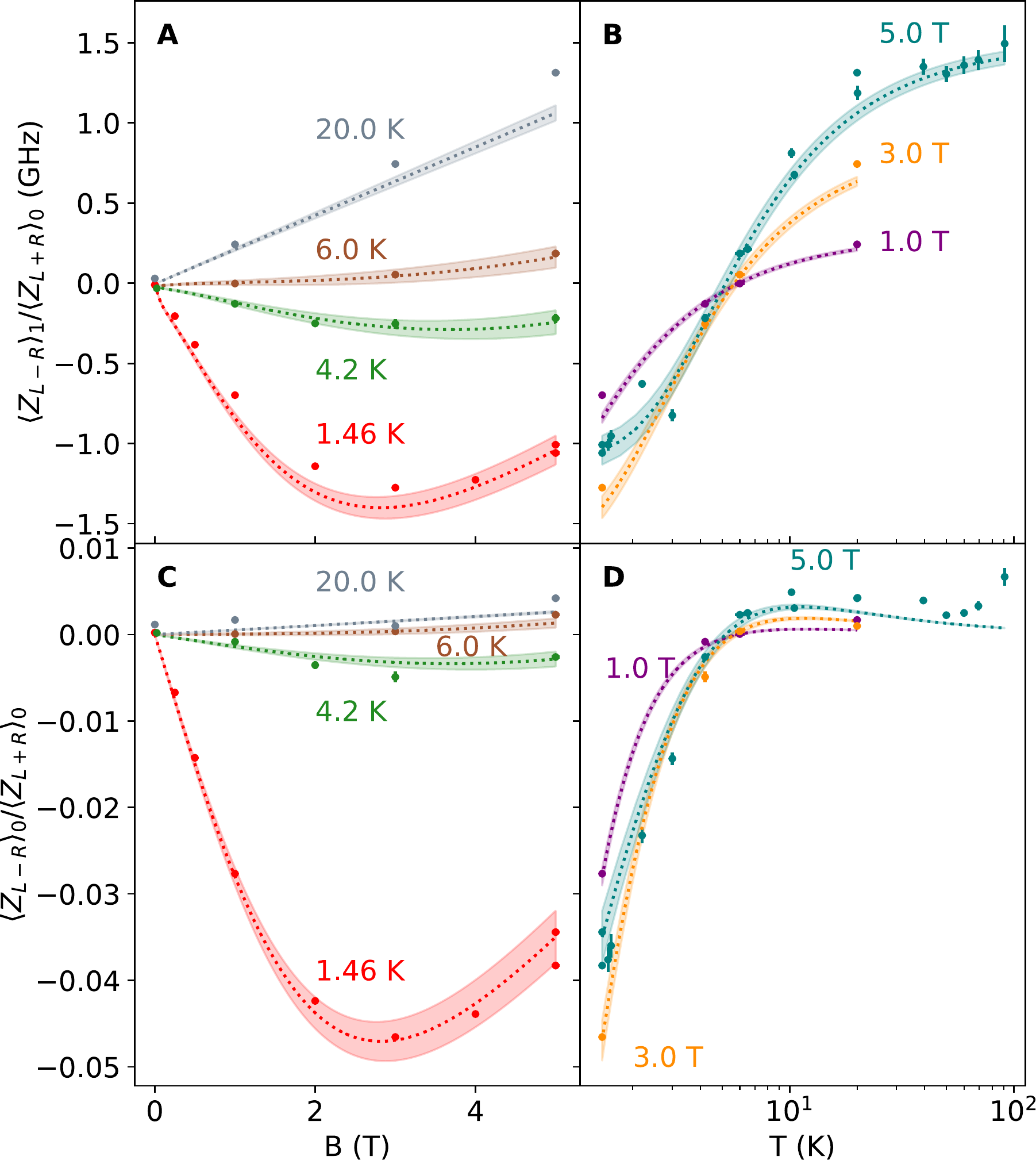}
	\caption{MCD values from (\ref{eqn-d}) and for varying magnetic field \textbf{(A)} and temperature \textbf{(B)}. MCD values from (\ref{eqn-Aratio}) for varying magnetic field \textbf{(C)} and temperature \textbf{(D)}. The points represent the values from curve fitting, the moment analysis points have been removed for clarity. The dashed curve is the Hamiltonian model using the average parameter values, $\lambda = \soparam \pm \soparamse $ GHz and $l = \lparam \pm \lparamse$, the shaded regions represent the $1\sigma$ uncertainty in the average parameter values.}
	\label{fig3}
\end{SCfigure*}

The results for a variety of temperatures and magnetic fields are shown in figure \ref{fig3}. The change in sign of the zeroth and first order moments for increasing temperature is due to the thermal populations of the ground state sub-levels changing, flipping the likelihood of absorbing a left or right circularly polarized photon.

\section{Analysis}

The data is fitted against the Hamiltonian (\ref{eqn-Hamiltonian-secular}) (via equations (\ref{eqn-Aratio}) and (\ref{eqn-d})) using only two free parameters, the spin-orbit coupling $\lambda$ and the orbital $g-$factor $l$, the other parameters $g =2$ and $\theta=\cos^{-1}\left(1/\sqrt{3}\right)$ were kept fixed. This fit obtains the values $\lambda = \soparam \pm \soparamse $ GHz and $l = \lparam \pm \lparamse$. Due to the difference in methods, the error analysis for both the curve fitting and the moment analysis was performed via a Monte-Carlo method. The raw spectra was modulated with random normally distributed noise with an amplitude equal to the standard deviation of the signal to the sides of the main feature. This was chosen as it represents the noise in the photocurrent signal for a particular spectra. The entire fitting procedure was re-run multiple times to obtain a distribution of the extracted parameters with an associated standard error and mean.

These results show that the orbital $g-$factor $l$ is reduced by an order of magnitude from the \nvm\, $^3E$ level, of which there are three published values (0.10(1) \cite{reddy1987two}, 0.22 \cite{HANZAWA1993137} and 0.164(6) \cite{Braukmann2018}). There is only one published value of the orbital $g-$factor for the \nvz\, \te\, state of 0.017(2).\cite{Braukmann2018} This is in reasonable agreement with our value of $l=\lparam(5)$. The spin-orbit parameter $\lambda$ is also reduced from its corresponding value in the \nvm\, $^3E$ level ($\lambda = 5.3$ GHz \cite{batalov2009low,TamaratManson2008}). However, the reduction is not as large as for the orbital $g-$factor (a factor of $\sim1/2$ rather than $\sim1/10$).

\section{Discussion}

There are a number of possible explanations for the differing reductions of the orbital $g-$factor and spin-orbit parameter from \nvm\, to \nvz. Both parameters are affected by the same Ham reduction factor \cite{FrankHam1965DJT_quenching,ham1968-epr-2e} arising from the JT interactions of the \nvz\, \te\, or \nvm\, $^3E$ levels. If the JT interaction of \nvz\, is larger than of \nvm\, (as indicated by features observed in piezospectroscopy of \nvz\,\cite{Davies_1979}), then this would explain the reduced size of these parameters in \nvz\,. However, this explanation would imply that both parameters should be reduced by the same factor. 

An alternate or complimentary explanation can be found in how the parameters depend on the defect's molecular orbital structure and local electrostatic potentials. The orbital $g-$factor is proportional to the reduced matrix element $\me{e(\vec{r};\vec{R})}{\lvert l_z\rvert}{e(\vec{r};\vec{R})}$, whereas the spin-orbit parameter is $\propto\me{e(\vec{r};\vec{R})}{\lvert [\vec{\nabla}V_{Ne}(\vec{r_i},\vec{R})\times \vec{p}]_z \rvert}{e(\vec{r};\vec{R})}$, where $l_z$ is the orbital angular momentum operator along the axis of the NV center, $[\vec{\nabla}V_{Ne}(\vec{r_i},\vec{R})\times \vec{p}]_z$ is similarly the axial component of the orbital operator of the spin-orbit interaction, $V_{Ne}$ is the electrostatic interaction between the defect electrons and the nuclei,  $\vec{r}$ are the electron coordinates, $\vec{R}$ are the nuclear coordinates, and $e(\vec{r};\vec{R})$ are $e$ molecular orbitals in the Born-Oppenheimer approximation \cite{doherty2011negatively}.

Owing to the different charge states of \nvz\, and \nvm, their nuclear coordinates and molecular orbitals differ \cite{doherty2013nitrogen,gali2009nv0}. Thus, the reduced matrix elements demonstrate that both the orbital $g-$factor and spin-orbit parameter will differ between the two charges states due to the molecular orbital differences, and the spin-orbit parameter will additionally differ due to differences in the electrostatic potential $V_{Ne}$. This additional dependence of the spin-orbit parameter over the orbital $g-$factor is the likely reason why the parameter reduction differ when comparing \nvz\, and \nvm.

Our observations do not immediately explain why fine-structure has not been observed in EPR or PLE measurements of the \nvz\, ground-state. Based on the spin-orbit parameter determined here and assuming the use of a 9.6 GHz X-band EPR spectrometer, the \nvz\, $^2E$ EPR features are expected to be at $B = \pm(\pm f-\lambda)/\mu_Bg=\pm0.16\,\text{T},\pm0.53\,\text{T}$ which are within the available field range of a typical X-band EPR spectrometer. Additionally, it has been shown that strong illumination with a green laser can photo-convert \nvm\, to \nvz. Thus, samples could be conveniently pumped to contain more \nvz\, centers \cite{manson2005photo,neil2018nv-n-pair-1b}. Additionally, a tunable high-resolution laser should also be able to see these features in PLE.

We believe that there are three major reasons why the EPR signals of the $^2E$ have not been previously observed: (1) A reduction in angular momentum could also be influenced by fast averaging over the orbital states by a weak JT coupling to a bath of acoustic $E$ modes. This is seen in the \nvm\, $^3E$ state at room temperatures \cite{rogers2009time,Tarus2015electron-phonon} as a removal of spin-orbit splittings. The remaining EPR signal would then be obscured by other spin$-1/2$ paramagnetic spins in the diamond sample, primarily the substitutional nitrogen or P1 center. (2) Another reason would be if there is a large-strain broadening of the resonances. For the limit of a strain distribution width ($\Gamma$) which is larger than spin-orbit splitting $(\Gamma \gg \lambda)$, only a central spin$-1/2$ resonance remains. This will be obscured for the same reason mentioned above. (3) The final reason is that measurements have simply not been performed in the correct spectral range or the signal has been overlooked.

\begin{SCfigure*}
	\centering
	\includegraphics[width=0.4\textwidth]{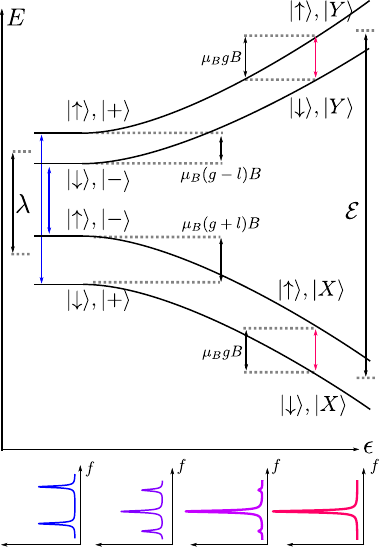}
	\caption{Behavior of \te\, spin-levels for increasing strain ($\epsilon$) in a non-zero magnetic field. In the high strain limit ($\mathcal{E} \gg \lambda$) the system is better represented by the $\left\{\ket{X}, \ket{Y}\right\}$ strain basis than the $\left\{\ket{+},\ket{-}\right\}$ orbital basis. The colored arrows represent the observable EPR resonances. Example EPR spectra for increasing strain are shown at the bottom of the figure, with increasing strain from left to right. The spectra changes from exhibiting spin-orbit, orbital Zeeman and spin Zeeman behavior to simple spin Zeeman behavior.}
	\label{fig4}
\end{SCfigure*}

We have modeled the electron-phonon interactions and have found them to produce a negligible effect on the \nvz\, spin resonances (see supplementary information), ruling out the first reason. For explanation 3, it seems unlikely that the signal was over looked, given targeted efforts to look for it \cite{PhysRevB.77.081201}, leaving explanation 2. We have modeled the second explanation (see supplementary information) and we find that a wide strain distribution significantly reduces the contribution of the strained centers to the total EPR spectrum. We model this by introducing a strain shift to the spin resonances $\mathcal{E}$, resulting in the spin-resonances (neglecting $l=\lparam$) at $\mathcal{B}$ and $\mathcal{B}\pm\Delta$ where $\mathcal{B} = g \mu_B B_z$ and $\Delta = \sqrt{\lambda^2 + \mathcal{E}^2}$. These spin resonances have the associated transition amplitudes $\cos^2\theta$ and $\sin^2\theta$, where $\theta = \tan\left(\mathcal{E}/\lambda\right)$. By constraining the oscillator strength to be conserved over the integrated spectral band, the intensity of the above resonances are $I_\mathcal{B} = \frac{\Gamma}{\Gamma+\lambda}$ and $I_\varepsilon = \frac{\lambda}{\Gamma + \lambda}$, where $\Gamma$ is the width of the strain distribution, here assumed to be Lorentzian. In the limit that $\Gamma \gg \lambda$, then $I_\varepsilon \rightarrow 0$ and $I_\mathcal{B} \rightarrow 1$, resulting in spectra with a single resonance only at the free spin$-1/2$ resonance frequency, this behavior is demonstrated in figure \ref{fig4}. The limit of $\Gamma \gg \lambda$ is reasonable, as the stress susceptibility and resulting width of the \nvz\, orbital electronic states are both of order THz \cite{Davies_1979}.

\section{Conclusion}
This work has determined the fine structure of the \nvz\, electronic states from low temperature and high magnetic field MCD measurements and found agreement by using simple spin-Hamiltonians to describe the ground and excited states. The spin-orbit interaction in the \nvz\, ground state $\lambda = \soparam \pm \soparamse $ GHz is found to be different to that of the \nvm\, excited state. This difference is attributed to the change electrostatic potential associated to variations in the nuclear coordinates between the two charge states. There is significant quenching of the angular momentum with an orbital $g-$factor of $l=\lparam \pm \lparamse$. We have discussed why measurement of \nvz\, ground-state fine structure has been elusive in previous EPR and PLE measurements and conclude that significant strain broadening of the $^2E$ resonances obscure the spin resonances.

\subsection*{}
We would like to acknowledge various funding sources for support during this work. In particular, NM would like to acknowledge the Australian Research Council through grants DP170103098. MD would like to acknowledge the Australian Research Council through grants DP170103098 and DE170100169. EK would like to acknowledge the Australian Research Council through grants DP110104565 and DP150103137.


\end{document}


\title{Supplementary Information: The ground-state fine-structure of the neutral nitrogen-vacancy defect in diamond}

\maketitle

\section{Transforming circularly polarized light to the NV coordinate frame}

During the MCD measurements, circularly polarized light was incident on a $\ms{100}$ face of the diamond sample. The NV centres are orientated along the $\ms{111}$ directions within the diamond sample. As such, the circularly polarized light must be transformed to the NV coordinate system. Defining a rotation that transforms the coordinate systems as defined in figure 1 in the main text gives the rotation matrix
\begin{align}
R = \left(
\begin{array}{ccc}
 1/\sqrt{6} & -1/\sqrt{2} & 1/\sqrt{3} \\
 1/\sqrt{6} & 1/\sqrt{2} & 1/\sqrt{3} \\
 -\sqrt{2/3} & 0 & 1/\sqrt{3}  
\end{array} \right).
\end{align}

As can been seen by the columns of the matrix, this transformation maps the coordinate vectors such that $R.[100] = [11\bar{2}],\;R.[010]=[\bar{1}10]$ and $R.[001]=[111]$. We define the right and left circularly polarized light electric field amplitudes in the crystal $[X,Y,Z]$ coordinate system as,
\begin{align}
	E_L &= \frac{1}{\sqrt{2}} \left( \begin{array}{c} 1 \\ i \\ 0 \end{array} \right) \\
	E_R &= \frac{1}{\sqrt{2}} \left( \begin{array}{c} 1 \\ -i \\ 0 \end{array} \right).
\end{align}
Applying the rotation operator gives the new left and right circularly polarized field amplitudes in the NV coordinate system $[x,y,z]$ as,
\begin{align}
	E_l &= \left(\begin{array}{c}
	-\frac{i}{2}+\frac{1}{2\sqrt{3}} \\
	\frac{i}{2}+\frac{1}{2\sqrt{3}} \\
	-\frac{1}{\sqrt{3}}
	\end{array}\right) \\
	E_r &= \left(\begin{array}{c}
	\frac{i}{2}+\frac{1}{2\sqrt{3}} \\
	-\frac{i}{2}+\frac{1}{2\sqrt{3}} \\
	-\frac{1}{\sqrt{3}}
	\end{array}\right).
\end{align}
The $E_z$ component is neglected as it doesn't drive any optical transitions of the NV centre. The resulting fields are elliptically polarized in the $xy$ plane with a phase lag between $E_x$ and $E_y$ of $\pm 2\pi/3$.

As described by Runciman and Manson \cite{Runciman_1988} the effect of this change in the light polarization introduces a modification to the MCD analysis. Since the left circularly polarised light in the lab frame is elliptical in the NV frame it contains both left and right circularly polarized components. Similarly, for the right circularly polarized light in the lab frame. This unwanted additional component reduces the differential absorption but not the total absorption. As such, the ratio of the spectral moments (and values from fitting lineshapes) requires a scaling of $\sqrt{3}/2$. Note that, Runciman and Manson scale their spectra by 1/2, however, they are also including the effect of the field misalignment on the Zeeman splitting ($\cos\theta = 1/\sqrt{3}$). This is included in our definition of the spin-Hamiltonian and is not required here.

\section{Electron-phonon broadening of the \nvz\, \te\, spin-resonances}

Here we present a model for electron-phonon broadening of the \nvz\, \te\, spin-resonances from first principles. This is used to explore the possibility of electron-phonon processes broadening and obscuring the appearance of \nvz\, spin-resonances from previous EPR measurements.

\subsection{Static strain interaction}

Applying group theory, the static strain interaction in the \nvz\, coordinate system can be immediately written down as

\begin{equation}
V_{\text{str}}=\left[\kappa _1\left(\epsilon _{\text{xx}}-\epsilon _{\text{yy}}\right)+2\kappa _2\epsilon _{\text{xz}}\right]L_z-\left[-2\kappa _1\epsilon
_{\text{xy}}+2\kappa _2\epsilon _{\text{yz}}\right]L_x\\
\\
=A_xL_z-A_yL_x
\label{eqn-strain-interaction-XY-basis}
\end{equation}

where \(\left.L_z=|X\right\rangle \langle X|-|Y\rangle \langle Y|\) and \(\left.L_x=|X\right\rangle \langle Y|+|Y\rangle \langle X|\) are the orbital
operators in the basis of strain eigenstates \(\{|X\rangle ,|Y\rangle \}\), \(\kappa _1\) and \(\kappa _2\) are the strain susceptibility parameters,
and \(\epsilon _{\text{ij}}\) are the strain components. Transforming to the orbital basis with well-defined angular momentum (ie spin-orbit/ magnetic
field eigenstates) \(|\pm \rangle =\mp (1/\surd 2)(|X\rangle \pm i|Y\rangle )\), the interaction becomes

\begin{equation}
 V_{\text{str}}=A\left(e^{{i\phi }}L_-+e^{-{i\phi }}L_+\right)
\end{equation}

where \(\left.L_{\pm }=|\pm \right\rangle \langle \mp |\), $A = \sqrt{A_x^2+A_y^2}$ and $\tan\phi = \frac{A_y}{A_x}$.

Davies \cite{davies1981jahnteller} performed uniaxial stress PL measurements of \nvz\, and determined the full set of stress susceptibility parameters. The relevant parameters are $b = -15$ cm$^{-1}$/GPa and $c = -15$ cm$^{-1}$/GPa which are in units of stress and defined with respect to the crystallographic coordinate system. Translating to units of strain in the NV coordinate system yields \cite{nanomechanical_sensing_barson}
\begin{align}
\kappa_1 &= -(-b(C_{11}-C_{12})-2c C_{44}) = -0.57~~\text{eV/strain}	\nonumber \\
\kappa_1 &= -(\sqrt{2}b(C_{11}-C_{12})-\sqrt{2}c C_{44})= 0.18~~\text{eV/strain}
\end{align}
where $[C_{11},C_{12},C_{44}] = [1076,135,576]$ GPa are the elements of the diamond stiffness tensor \cite{kaxiras2003atomic}.

\subsection{Interactions with acoustic phonons}
	
The interaction of the \(\text{NV}^0\)with long-wavelength phonons can be approximated via a dynamic strain interaction as long as the strain gradients
are insignificant over the dimension of the \(\text{NV}^0\) electronic orbitals. In this case, the electron-phonon interaction expressed in the strain
eigenbasis and angular momentum eigenbasis are, respectively

\begin{align}
V_{e-p}&=\sum _{\overset{\rightharpoonup }{k},p}\left(A_{\overset{\rightharpoonup }{k},p,x}L_z-A_{\overset{\rightharpoonup }{k},p,y}L_x\right)\left(\frac{\hbar
}{2{\rho V\omega }_{{kp}}}\right){}^{1/2}\left(a_{\overset{\rightharpoonup }{k},p}+a_{\overset{\rightharpoonup }{k},p}^{\dagger }\right) \nonumber \\
=&\sum _{\overset{\rightharpoonup }{k},p}A_{\overset{\rightharpoonup }{k},p}\left(e^{{i\phi }_{\overset{\rightharpoonup }{k},p}}L_-+e^{-{i\phi
		}_{\overset{\rightharpoonup }{k},p}}L_+\right)\left(\frac{\hbar }{2{\rho V\omega }_{{kp}}}\right){}^{1/2}\left(a_{\overset{\rightharpoonup
	}{k},p}+a_{\overset{\rightharpoonup }{k},p}^{\dagger }\right)
\end{align}

where \(a_{\overset{\rightharpoonup }{k},p}\) and \(a_{\overset{\rightharpoonup }{k},p}^{\dagger }\) are the annihilation and creation operators
of the phonon mode with wavevector \(\overset{\rightharpoonup }{k}\), polarization \(p=t_1,t_2,l\), frequency \(\omega _{{kp}}={kc}_p\),
and normalised displacement vector \(\overset{\rightharpoonup }{u}_{\overset{\rightharpoonup }{k},p}\), \(c_{p }=c_t,c_l\) are the transverse and
longitudinal acoustic velocities of diamond, \(V\) is volume of the diamond unit cell, \(\rho\) is the density of diamond, and \(A_{\overset{\rightharpoonup
	}{k},p,\pm }\) and \(\phi _{\overset{\rightharpoonup }{k},p}\) are defined by the strain \(\epsilon _{\text{ij}}\left(\overset{\rightharpoonup }{k},p\right)=(1/2)\left(\partial
_iu_{\overset{\rightharpoonup }{k},p,j}+\partial _ju_{\overset{\rightharpoonup }{k},p,i}\right)\) induced per unit excitation of the phonon mode.
The displacement vector is normalized via \(\int _V\overset{\rightharpoonup }{u}_{\overset{\rightharpoonup }{k},p}^*\cdot \overset{\rightharpoonup
}{u}_{\overset{\rightharpoonup }{k},p}d^3r=V\). 

The interaction has been expressed in the different eigenbases so that the correct basis can be chosen when applying time-dependent perturbation
theory to calculate the electron-phonon scattering rates. In the low strain regime, where the strain splitting \(\hbar \Delta _{\epsilon }\) is much
smaller than the spin-orbit splitting \(\hbar \Delta _{\lambda }\), the stationary orbital states will be best described by the angular momentum
eigenbasis. Whilst in the high strain regime \(\hbar \Delta _{\epsilon }>>\hbar \Delta _{\lambda }\), the strain eigenbasis will be the best description.

\subsection{Electron-phonon scattering in the low strain regime (adapted from ref. \cite{1367-2630-17-4-043011})}

\subsubsection{One-phonon processes}

Applying first-order perturbation theory, the rate \(\Gamma _{-+}^{(1)}\) of the transition \(|-\rangle \to |+\rangle\) via the absorption of a phonon
and the rate \(\Gamma _{+-}^{(1)}\) of the opposite transition via the emission of a phonon are

\begin{align}
\Gamma _{-+}^{(1)}&=\frac{2\pi }{\hbar ^2}\sum _{\overset{\rightharpoonup }{k},p}\frac{\hbar }{2\text{$\rho $V$\omega $}_{\text{kp}}}|A_{\overset{\rightharpoonup
	}{k},p}|^2n(\Delta )\delta \left(\Delta -\omega _{\text{kp}}\right)=\frac{2\pi }{\hbar ^2}\eta \Delta ^3n(\Delta ) \nonumber \\
\Gamma _{+-}^{(1)}&=\frac{2\pi }{\hbar ^2}\sum _{\overset{\rightharpoonup }{k},p}\frac{\hbar }{2\text{$\rho $V$\omega $}_{\text{kp}}}|A_{\overset{\rightharpoonup
	}{k},p}|^2\left[n\left(\omega _{\text{kp}}\right)+1\right]\delta \left(\Delta -\omega _{\text{kp}}\right)=\frac{2\pi }{\hbar ^2}\eta \Delta ^3[n(\Delta
)+1]
\end{align}

where \(\hbar \Delta =E_+-E_-\) is the splitting of the orbital states, \(n(\omega )=\left(e^{\hbar \omega \left/k_B\right.T}-1\right){}^{-1}\) is
the Bose-Einstein population of phonon modes of frequency $\omega $ in thermal equilibrium, and

\begin{align}
\eta &=\omega ^{-3}\sum _{\overset{\rightharpoonup }{k},p}\frac{\hbar }{2\text{$\rho $V$\omega $}_{\text{kp}}}|A_{\overset{\rightharpoonup }{k},p}|^2\delta
\left(\omega -\omega _{\text{kp}}\right) \nonumber \\
&=\omega ^{-3}\sum _p\frac{V}{(2\pi )^3}\int d^3k\frac{\hbar }{2\text{$\rho $V$\omega $}_{\text{kp}}}|A_{\overset{\rightharpoonup }{k},p}|^2\delta
\left(\omega -\omega _{\text{kp}}\right)
 \nonumber \\
&=\frac{\hbar \left(3c_l^5+2c_t^5\right)\left(\kappa _1^2+\kappa _2^2\right)}{15c_l^5c_t^5\pi ^2\rho }
\end{align}

If the temperature is such that \(k_BT>>\hbar \Delta\), the rates become approximately linear in temperature
\begin{equation}
\Gamma _{+-}^{(1)}\approx e^{\hbar \Delta \left/k_B\right.T}\Gamma _{-+}^{(1)}\approx \frac{2\pi }{\hbar ^2}\eta \Delta ^2\frac{k_B}{\hbar }T
\label{eqn-W1mp}
\end{equation}

\begin{figure}[h]
	\centering
	\includegraphics[width=0.6\textwidth]{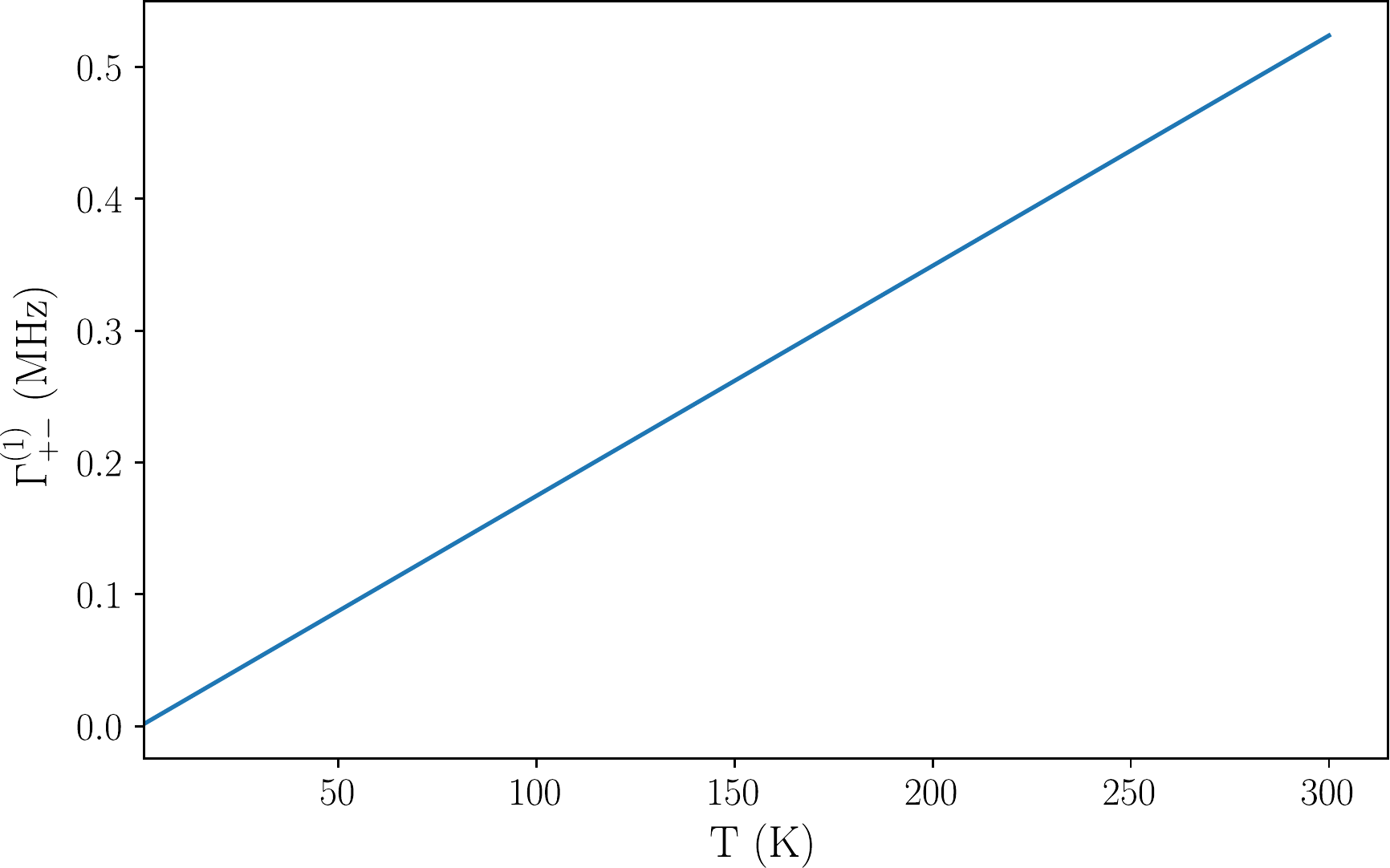}
	\caption{Calculated one-phonon transition rate between the orbitally split \te\, states.}
	\label{fig-supp1}
\end{figure}

Using the parameters $c_l$ and $c_t$ of 18146 m/s and 12354 m/s respectively, $\rho$  of 3512 kg/m$^3$ and a zero-field orbital splitting of $2\lambda = 5.25$ GHz, the one-phonon transition rate expressed in equation (\ref{eqn-W1mp}) can be numerically evaluated and is shown in figure \ref{fig-supp1}.

\subsubsection{Two-phonon processes}

Since the electron-phonon interaction in the angular momentum eigenbasis only contains off-diagonal orbital matrix elements, the initial and final
orbital states cannot differ in all two-phonon processes. In other words, the processes cannot induce an orbital transition, only a dephasing of
the orbital state. The only such processes that are also allowed by energy conservation are Raman-type processes, in which phonons of the same frequency
are absorbed and emitted (two-phonon elastic scattering).

Applying second-order perturbation theory, the scattering rates are

\begin{align}
\Gamma _{\text{++}}^{(2)}&=\Gamma _{\text{--}}^{(2)}
=\frac{2\pi }{\hbar ^2}\sum _{\overset{\rightharpoonup }{k},p}\sum _{\overset{\rightharpoonup }{k}',p'}\frac{\hbar }{2\text{$\rho $V$\omega $}_{\text{kp}}}\frac{\hbar
}{2\text{$\rho $V$\omega $}_{k'p'}}\left|A_{\overset{\rightharpoonup }{k},p}|^2\right|A_{\overset{\rightharpoonup }{k}',p'}|^2|\frac{1}{\Delta -\omega
	_{\text{kp}}}+\frac{1}{\Delta +\omega _{k'p'}}|^2n\left(\omega _{\text{kp}}\right)\left[n\left(\omega _{k'p'}\right)+1\right]\delta \left(\omega
_{\text{kp}}-\omega _{k'p'}\right) \nonumber
\\
&=\frac{2\pi }{\hbar ^2}\int _0^{\Omega }\eta ^2\omega ^6|\frac{1}{\Delta -\omega }+\frac{1}{\Delta +\omega }|^2n(\omega )[n(\omega )+1]\text{d$\omega
	$}
\end{align}

where \(\hbar \Omega \sim 70\) meV is the acoustic phonon cutoff frequency of diamond. If the temperature is such that the dominant contribution to the integral comes from phonons with frequencies much larger than $\Delta $, then it can be approximated by lowest order term in the absolute square

\begin{equation}
\Gamma _{\text{++}}^{(2)}=\Gamma _{\text{--}}^{(2)}\\
\\
\approx \frac{2\pi }{\hbar ^2}\int _0^{\Omega }\eta ^2\omega ^6\left(\frac{4\Delta ^2}{\omega ^4}\right)n(\omega )[n(\omega )+1]\text{d$\omega $}\\
\\
=\frac{8\pi }{\hbar ^2}\eta ^2\Delta ^2\int _0^{\Omega }\omega ^2n(\omega )[n(\omega )+1]\text{d$\omega $}
\end{equation}

Making the replacements \(x=\hbar \omega \left/k_B\right.T\) and \(X_c=\hbar \Omega \left/k_B\right.T\), this becomes

\begin{equation}
\Gamma _{\text{++}}^{(2)}=\Gamma _{\text{--}}^{(2)}\approx \frac{8\pi }{\hbar ^2}\eta ^2\Delta ^2\frac{k_B^3T^3}{\hbar ^3}I_3(T)
\end{equation}

where

\begin{equation}
I_3(T)=\int _0^{X_c}x^2\frac{e^x}{\left(e^x-1\right)^2}\text{dx}
\end{equation}

thereby revealing a \(T^3\) dependence in the limit \(X_c\to \infty\).

\begin{figure}[h]
	\centering
	\includegraphics[width=0.6\textwidth]{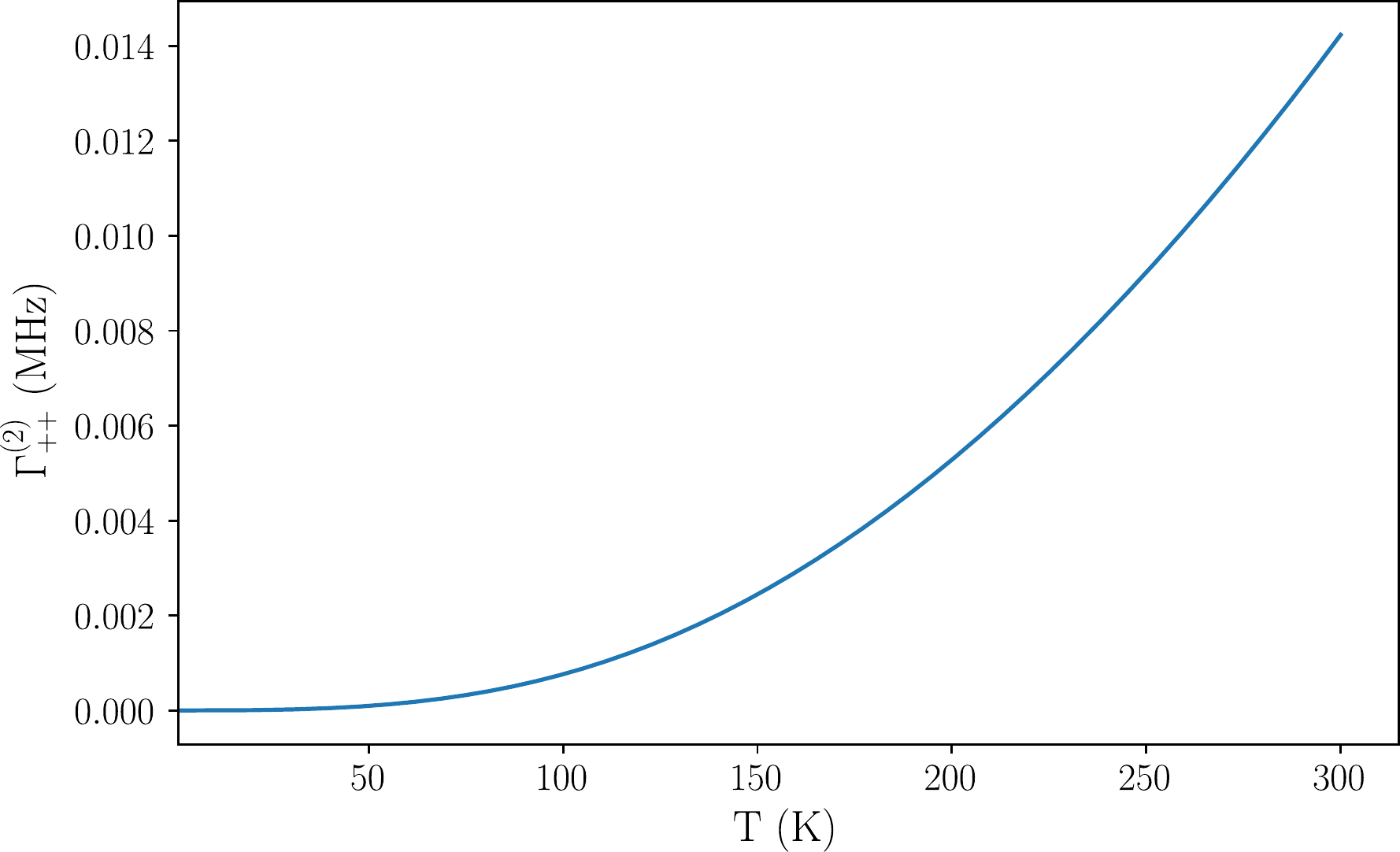}
	\caption{Calculated two-phonon dephasing rate between the orbitally split \te\, states.}
	\label{fig-supp2}
\end{figure}

Using the same parameters as for the one-phonon transition rate the two-phonon dephasing rate can be calculated and is shown in figure \ref{fig-supp2}

\subsection{Electron-phonon scattering in the high strain regime}

The following modelling draws heavily from Ref. \cite{Tarus2015electron-phonon}. In the high strain regime the orbital quantum numbers are better described in the $\ket{X},\ket{Y}$ strain-basis than the orbital-basis $\ket{+},\ket{-}$.

\subsubsection{One-phonon processes}

 Following the same methodology using the strain eigenbasis and defining \(\hbar \Delta =E_x-E_y\) in this case, we find the first-order rates to be

\begin{align}
\Gamma _{\text{yx}}^{(1)}&=\frac{1}{2}\Gamma _{-+}^{(1)} \nonumber \\
\Gamma _{\text{xy}}^{(1)}&=e^{\hbar \Delta \left/k_B\right.T}\Gamma _{\text{yx}}^{(1)}.
\end{align}

As such, these values were presented in the previous section.

\subsubsection{Two-phonon processes}

Likewise applying the same methodology, the elastic two-phonon processes leading to pure dephasing are

\begin{equation}
\Gamma _{\text{xx}}^{(2)}=\Gamma _{\text{yy}}^{(2)}=\frac{1}{4}\Gamma _{\text{++}}^{(2)}
\end{equation}

In the strain eigenbasis, Raman-type two-phonon processes may also drive transitions between the orbital states

\begin{align}
\Gamma _{\text{yx}}^{(2)}=\frac{2\pi }{\hbar ^2}\int _0^{\Omega }\eta ^2\omega ^4\left(1+\frac{\Delta }{\omega }\right)^3\left(1+\frac{1}{(1+\Delta
	/\omega )^2}\right)n(\omega +\Delta )[n(\omega )+1]\text{d$\omega $}\nonumber \\
=\frac{2\pi }{\hbar ^2}\int _0^{\Omega }\eta ^2\omega ^4\left(\left(1+\frac{\Delta }{\omega }\right)^3+1+\frac{\Delta }{\omega }\right)n(\omega +\Delta
)[n(\omega )+1]\text{d$\omega $}\nonumber \\
\approx \frac{2\pi }{\hbar ^2}\int _0^{\Omega }\eta ^2\omega ^4\left(2+4\frac{\Delta }{\omega }+3\frac{\Delta ^2}{\omega ^2}+\frac{\Delta ^3}{\omega
	^3}\right)n(\omega +\Delta )[n(\omega )+1]\text{d$\omega $}
\end{align}

Making the replacements \(x=\hbar \omega \left/k_B\right.T\), \(x_{\Omega }=\hbar \Omega \left/k_B\right.T\) and \(x_{\Delta }=\hbar \Delta \left/k_B\right.T\),
this becomes

\begin{equation}
\Gamma _{\text{yx}}^{(2)}\approx \frac{4\pi }{\hbar ^2}\eta ^2\frac{k_B^5T^5}{\hbar ^5}I_5(T)
\end{equation}

where

\begin{equation}
I_5(T)=\int _0^{x_{\Omega }}\left(x^4+2x_{\Delta }x^3+\frac{3}{2}x_{\Delta }^2x^2+\frac{1}{2}x_{\Delta }^3x\right)\frac{e^x}{\left(e^x-1\right)\left(e^{x+x_{\Delta
	}}-1\right)}\text{dx}
\end{equation}

Via the principle of detailed balance, the other rate is

\begin{equation}
\Gamma _{\text{xy}}^{(2)}=e^{\hbar \Delta \left/k_B\right.T}\Gamma _{\text{yx}}^{(2)}
\end{equation}

Note that the two-phonon absorption of and two-phonon emission processes have been ignored because owing to the \(\sim \omega\) scaling of the electron-phonon coupling and \(\sim \omega ^2\) scaling of the phonon density of modes, they are expected to be much slower than the one-phonon and two-phonon Raman processes that involve higher frequency modes.

\begin{figure}[h]
	\centering
	\includegraphics[width=0.6\textwidth]{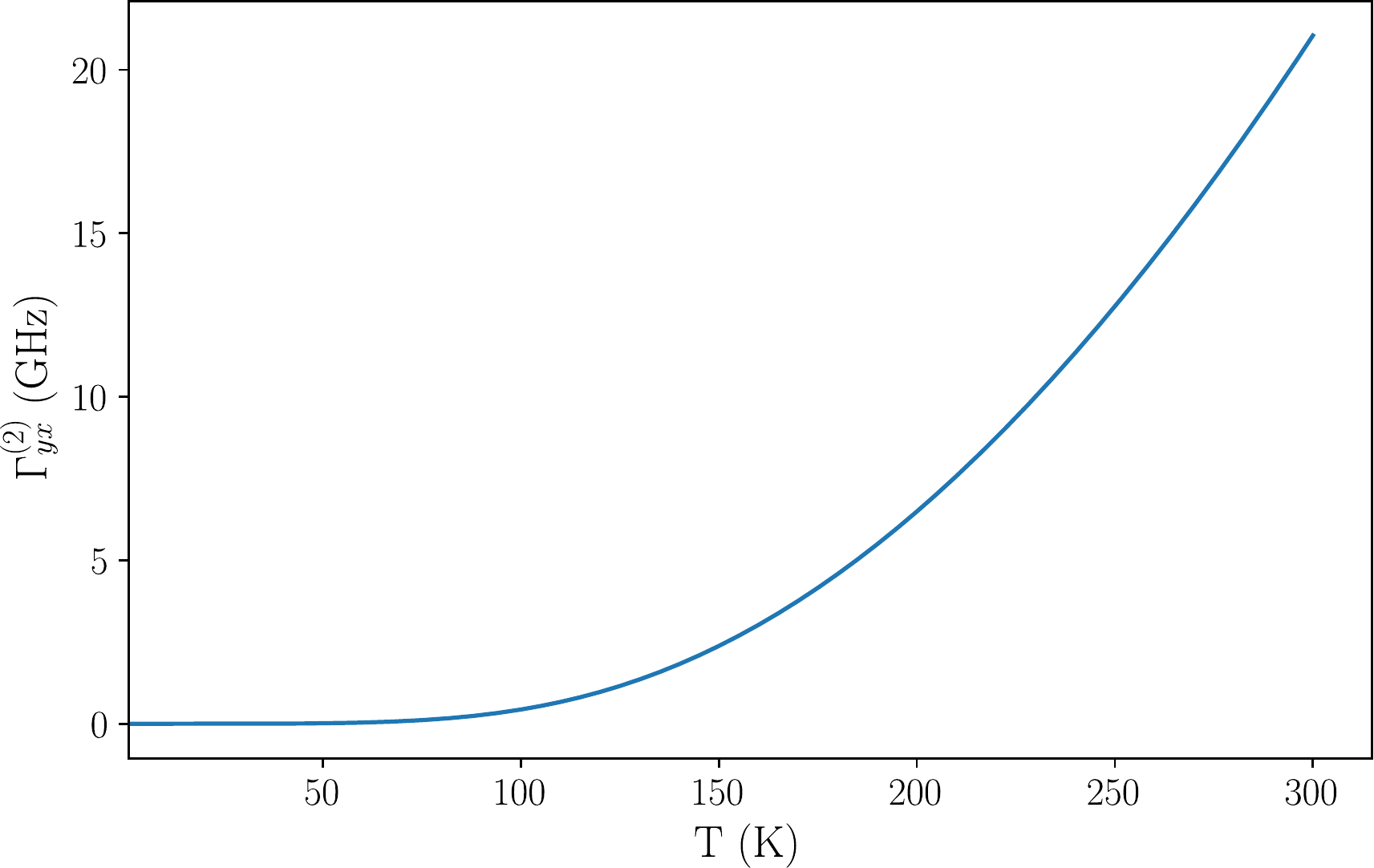}
	\caption{Calculated two-phonon dephasing rate between the orbitally split \te\, states.}
	\label{fig-supp3}
\end{figure}

\subsection{Motional narrowing of the \(\text{NV}^0\) ground state spin resonance}

Drawing from Ref. \cite{Tarus2015electron-phonon}, the lineshape of the ground state spin resonance is

\begin{equation}
I(\omega )=\text{Re}\left\{\frac{1}{\pi }\int _0^{\infty }\phi (t)e^{\text{i$\omega $t}}\text{dt}\right\}
\end{equation}

where

\begin{align}
\phi (t)&=e^{-\text{i$\omega $}_st}e^{-Kt/2}\left[\cosh (\mathcal{E}t)+\frac{\alpha }{\mathcal{E}}\sinh (\mathcal{E}t)\right]\\
\mathcal{E}&=\left[\frac{K^2}{4}-\frac{\Delta _s^2}{4}-\frac{i}{2}{PK\Delta}_s\right]{}^{1/2}\\
\alpha &=\frac{K}{2}-\frac{i}{2}{P\Delta }_s\\
P&=\frac{e^{-x_{\Delta }}-1}{e^{-x_{\Delta }}+1}
\end{align}

\(K\) is the net transition rate between the orbital states, \(\Delta _s\) is the change in the electron spin resonance when the orbital state changes
(for an aligned magnetic field or sufficiently small magnetic field \(\Delta _s=\Delta _{\lambda }\)), and \(\omega _s=\gamma _eB\) is the spin resonance
frequency in the absence of any orbital coupling. Note that \(P\) is the difference in population between the two orbital states in thermal equilibrium.
Like \(\Delta\), { }\(K\) depends on whether the center is in the low or high strain regime

\begin{equation}
\Delta =\left\{
\begin{array}{cc}
\Delta _{\lambda } & \Delta _{\lambda }>>\Delta _{\epsilon } \\
\Delta _{\epsilon } & \Delta _{\epsilon }>>\Delta _{\lambda } \\
\end{array}
\right.
\end{equation}

\begin{equation}
K=\left\{
\begin{array}{cc}
\Gamma _{-+}^{(1)}\left(1+e^{x_{\Delta }}\right) & \Delta _{\lambda }>>\Delta _{\epsilon } \\
\left(\Gamma _{\text{yx}}^{(1)}+\Gamma _{\text{yx}}^{(2)}\right)\left(1+e^{x_{\Delta }}\right) & \Delta _{\epsilon }>>\Delta _{\lambda } \\
\end{array}
\right.
\end{equation}

Note that we've ignored the orbital magnetic interaction since it is assumed that the magnetic field is sufficiently low that it is not comparable to spin-orbit.

\subsubsection{Fast exchange limit}

In the fast exchange limit \(K>>\Delta _s\)

\begin{equation}
\phi (t)\approx e^{-i\left(\omega _s+\left.\text{P$\Delta $}_s\right/2\right)t}e^{-\Gamma _{\text{MN}}t/2}
\end{equation}

which corresponds to a single Lorentzian centered at frequency \(\omega _S+\left.{P\Delta }_s\right/2\) and with a motionally narrowed FWHM
(in rad/s)

\begin{equation}
\Gamma _{\text{MN}}=\frac{2\Delta _s^2}{K}\frac{e^{-x_{\Delta }}}{\left(e^{-x_{\Delta }}+1\right){}^2}
\end{equation}

As expected, \(\Gamma _{\text{MN}}\) is inversely proportional to \(K\) in the fast exchange limit.

\subsubsection{Slow exchange limit}

In the slow exchange limit \(K<<\Delta _s\) { }

\begin{equation}
\phi (t)=\frac{1}{2}e^{-\text{i$\omega $}_st}e^{-\Gamma _{\text{MB}}t/2}\left[(1-P)e^{\text{i$\Delta $}_st/2}+(1+P)e^{-\text{i$\Delta $}_st/2}\right]
\end{equation}

which corresponds to two Lorentzians at frequencies \(\omega _S\pm \left.\Delta _s\right/2\), with relative amplitudes determined by the population
difference \(P\) and with a motionally broadened FWHM (in rad/s)

\begin{equation}
\Gamma _{\text{MB}}=K+2D
\end{equation}

where

\begin{equation}
D=\left\{
\begin{array}{cc}
\Gamma _{\text{++}}^{(2)}+\Gamma _{\text{--}}^{(2)} & \Delta _{\lambda }>>\Delta _{\epsilon } \\
\Gamma _{\text{xx}}^{(2)}+\Gamma _{\text{yy}}^{(2)} & \Delta _{\epsilon }>>\Delta _{\lambda } \\
\end{array}
\right.
\end{equation}

is the total pure dephasing rate. As expected, \(\Gamma _{\text{MB}}\) is proprotional to \(K\) in the slow exchange limit.\\

\subsection{Conclusion}

In the low strain limit, both the first order transition rate ($\Gamma_{+-}^{(1)}$) and second order dephasing rate ($\Gamma_{++}^{(2)}$) are slower than 1 MHz which is much smaller than the spin-orbit coupling $\lambda \sim$ GHz, as shown in figures \ref{fig-supp1} and \ref{fig-supp2}. In this slow exchange limit the EPR resonances will be motionally broadened to give a linewidth proportional to the electron-phonon rates. However, these rates are very slow and don't provide sufficient broadening to obscure the EPR resonances. \\

In the high strain limit, the two-phonon dephasing and transition rates are similar in magnitude to spin-orbit coupling at high temperature, as shown in figure \ref{fig-supp3}. As such, the system will only just start reaching the fast-exchange limit at room temperature and is otherwise in the intermediate or slow exchange limit. In this slow exchange limit the linewidth is motionally broadened and is proportional to $\Gamma_{xx}^{(2)}$. Which is very slow/narrow (see figure \ref{fig-supp3}). As such, this also doesn't provide sufficient broadening to obscure the EPR resonances. \\

Many EPR measurements in diamond have been performed at low temperature, as pointed out in the review article by Loubser and van Wyk \cite{loubser1978electron}. Additionally, the EPR measurements of the \nvz\, $^4A_2$ meta-stable state were also performed at low temperature \cite{PhysRevB.77.081201}. Since many of these measurements would be in the slow exchange limit, electron-phonon broadening processes must be ruled out as the reason why the \nvz\, \te\, spin resonances have been obscured from previous EPR measurements.

\section{Stress-broadening of the \nvz\, \te\, spin-resonances}

\subsection{Strain broadening of the NV0 spin resonances}

Adding the strain interaction from the previous section to the spin-Hamiltonian from the main text gives
\begin{align}
	H = g \mu_B S_z B_z + l \mu_B {L_z} B_z + \lambda {L_z} {S_z} +  \mathcal{E}_- L_+ + \mathcal{E}_- L_-,
\end{align}
where $\mathcal{E}_\pm = \mathcal{E}e^{\pm i \phi}$, $\phi=\tan\frac{A_y}{A_x}$ and $\mathcal{E} = \sqrt{A_x^2+A_y^2}$ where $A_x$ and $A_y$ are defined as in equation (\ref{eqn-strain-interaction-XY-basis}).
For simplicity we have assumed that the magnetic field is aligned along $S_z$ direction and made the substitutions $g\mu_B B_z \rightarrow \mathcal{B}$, $l/g\rightarrow l$ and $\mathcal{E}_\pm \rightarrow \mathcal{E}_\pm/4$ giving the Hamiltonian
\begin{align}
H = \mathcal{B} S_z + l \mathcal{B}L_z + \lambda {L_z} {S_z} +  (\mathcal{E}_- L_+ + \mathcal{E}_- L_-)/4.
\end{align}
Which has the resulting eigensolutions,
\begin{center}
	\begin{tabular}{|c|c|}
		\hline
		eigenfrequency & eigenstate \\ \hline
		$1/2(\Delta_+ + \mathcal{B})$ & $\ket{4} = \ket{a_+}\ket{\uparrow}$ \\ 
		$1/2(\Delta_- - \mathcal{B})$ & $\ket{3} = \ket{b_-}\ket{\downarrow}$ \\ 
		$1/2(-\Delta_+ + \mathcal{B})$ & $\ket{2} = \ket{b_+}\ket{\uparrow}$ \\ 
		$1/2(-\Delta_- - \mathcal{B})$ & $\ket{1} = \ket{a_-}\ket{\downarrow}$ \\
		\hline
	\end{tabular}
\end{center}
where $\Delta_\pm=\sqrt{\left(l\mathcal{B} \pm \lambda\right)^2 + \mathcal{E}^2}$ and the mixed angular momentum orbital states are $\ket{a_\pm} = \cos\frac{\theta_+\pm}{2}\ket{+} + \sin\frac{\theta_+\pm}{2}\ket{-}$ and $\ket{b_\pm} = -\sin\frac{\theta_+\pm}{2}\ket{+} + \cos\frac{\theta_+\pm}{2}\ket{-}$, with $\tan\theta_\pm = \frac{\mathcal{E}}{l\mathcal{B}\pm\lambda}$. We have simplified the solutions by assuming there is no phase difference between the $x$ and $y$ strain interactions i.e. $\phi=0$. The resulting spin resonance amplitudes and frequencies are thus
\begin{center}
	\begin{tabular}{|c|c|c|}
		\hline
		Transition & Energy & Amplitude \\ \hline
		 $1 \rightarrow 2$ & $\mathcal{B}-\frac{1}{2}(\Delta_+-\Delta_-)$ & $\left(\cos\frac{\theta_+}{2}\sin\frac{\theta_-}{2}-\cos\frac{\theta_-}{2}\sin\frac{\theta_+}{2}\right)^2$ \\
		 $1 \rightarrow 4$ & $\mathcal{B}+\frac{1}{2}(\Delta_++\Delta_-)$ & $\left(\cos\frac{\theta_+}{2}\cos\frac{\theta_-}{2}+\sin\frac{\theta_-}{2}\sin\frac{\theta_+}{2}\right)^2$ \\
		 $2 \rightarrow 3$ & $-\mathcal{B}+\frac{1}{2}(\Delta_++\Delta_-)$ & $\left(\cos\frac{\theta_+}{2}\cos\frac{\theta_-}{2}+\cos\frac{\theta_-}{2}\cos\frac{\theta_+}{2}\right)^2$ \\
		 $3 \rightarrow 4$ & $\mathcal{B}+\frac{1}{2}(\Delta_+-\Delta_-)$ & $\left(\cos\frac{\theta_+}{2}\sin\frac{\theta_-}{2}-\cos\frac{\theta_-}{2}\sin\frac{\theta_+}{2}\right)^2$\\ \hline
	\end{tabular}
\end{center}

If we make the simplification that \(l\sim 0\), then \(\Delta _+\sim \Delta _-\sim \Delta\) and \(\theta _+\sim -\theta _-\sim \theta\), and the
above becomes
\begin{center}
	\begin{tabular}{|c|c|c|}
		\hline
		Transition & Energy & Amplitude \\ \hline
		$1\rightarrow2$ & $\mathcal{B}$ & $\sin ^2\theta$  \\
		$1\rightarrow4$ & $\Delta +\mathcal{B}$ & $\cos ^2\theta$  \\
		$2\rightarrow3$ & $\Delta -\mathcal{B}$ & $\cos ^2\theta$  \\
		$3\rightarrow4$ & $\mathcal{B}$ & $\sin ^2\theta$  \\ \hline
	\end{tabular}
\end{center}

In the low strain limit \(\Delta \to \lambda\) and \(\theta \to 0\), meaning that we have two spin resonances with energy \(\Delta \pm \mathcal{B}\).
In the high strain limit \(\Delta \to \mathcal{E}\) and \(\theta \to  \pi /2\), meaning that we have a single spin resonance with energy \(\mathcal{B}\).
In the intermediate regime, we have multiple spin resonances.

In the simplest picture, the spin resonance spectrum of a single center as a function of strain is

\begin{equation}
f(\omega ,\mathcal{E})=2\sin ^2\theta  \delta (\omega -\mathcal{B})+\cos ^2\theta [\delta (\omega -\Delta -\mathcal{B})+\delta (\omega -\Delta +\mathcal{B})]
\end{equation}

where we've ignored the different thermal occupations of the states, and sources of broadening.

The spin resonance spectrum of an ensemble is the statistical average of the above over the strain distribution 
$D(\mathcal{E})$.

The spectrum is thus

\begin{equation}
F(\omega )=2\delta (\omega -\mathcal{B})\int _{-\infty }^{\infty }D(\mathcal{E})\sin ^2\theta  \text{d$\mathcal{E}$}+\int _{-\infty }^{\infty }D(\mathcal{E})\cos
^2\theta [\delta (\omega -\Delta -\mathcal{B})+\delta (\omega -\Delta +\mathcal{B})]\text{d$\mathcal{E}$}
\end{equation}

Applying the delta function composition property, where the delta function may be composed with a smooth function $g(x)$,
\begin{align}
\delta\left(g(x)\right) = \sum_i \frac{\delta(x-x_i}{\left|g'(x_i)\right|},
\end{align}
where the sum extends over all roots $x_i$ of $g(x)$. For example,
\begin{align}
\delta\left(x^2-\alpha^2\right) = \frac{1}{2|\alpha|} \left( \delta(x+\alpha) + \delta(x-\alpha)\right)
\end{align}

The above comes

\begin{align}
F(\omega )&=2\delta (\omega -\mathcal{B})\int _{-\infty }^{\infty }D(\mathcal{E})\sin ^2\theta  \text{d$\mathcal{E}$}+\int _{-\infty }^{\infty }D(\mathcal{E})\cos
^2\theta [\delta (\omega -\Delta -\mathcal{B})+\delta (\omega -\Delta +\mathcal{B})]\text{d$\mathcal{E}$}\nonumber
\\
&=2\delta (\omega -\mathcal{B})\int _{-\infty }^{\infty }D(\mathcal{E})\sin ^2\theta  \text{d$\mathcal{E}$}+ \cdots \\ \nonumber 
&\int _{-\infty }^{\infty }D(\mathcal{E})\frac{\lambda
	^2}{\mathcal{E}^2+\lambda ^2}\frac{1}{2}\left\{\left|\frac{\epsilon _-}{\left[\epsilon _-+\lambda ^2\right]{}^{1/2}}|^{-1}\left[\delta \left(\mathcal{E}-\epsilon
_-\right)+\delta \left(\mathcal{E}+\epsilon _-\right)\right]+\right|\frac{\epsilon _+}{\left[\epsilon _++\lambda ^2\right]{}^{1/2}}|^{-1}\left[\delta
\left(\mathcal{E}-\epsilon _-\right)+\delta \left(\mathcal{E}+\epsilon _-\right)\right]\right\}\text{d$\mathcal{E}$} \nonumber \\
&=2\delta (\omega -\mathcal{B})\int _{-\infty }^{\infty }D(\mathcal{E})\sin ^2\theta  \text{d$\mathcal{E}$}+\frac{\lambda ^2}{\left|\epsilon _-\right|\left[\epsilon
	_-^2+\lambda ^2\right]{}^{1/2}}D\left(\epsilon _-\right)+\frac{\lambda ^2}{\left|\epsilon _+\right|\left[\epsilon _+^2+\lambda ^2\right]{}^{1/2}}D\left(\epsilon
_+\right)
\end{align}

where \(\epsilon _{\pm }=\left[(\omega \pm \mathcal{B})^2-\lambda ^2\right]^{1/2}\). The second and third terms represent lines that diverge at \(\epsilon
_{\pm }=0\). This is avoided in reality via the intrinsic linewidth of the spin resonances due to dephasing/ relaxation. Mathematically, this must
be introduced through the convolution of the above with the intrinsic spin resonance lineshape.

What is important for the assessment of whether these lines are observable in EPR is the integrated areas of the lines (ie their oscillator strength). As each \nvz\, has two possible transitions, the integral of the above over all frequency must equal 2. Thus

\begin{equation}
2\int _{-\infty }^{\infty }\delta (\omega -\mathcal{B})\text{d$\omega $}\int _{-\infty }^{\infty }D(\mathcal{E})\sin ^2\theta  \text{d$\mathcal{E}$}+\int
_{-\infty }^{\infty }\frac{\lambda ^2}{\left|\epsilon _-\right|\left[\epsilon _-^2+\lambda ^2\right]{}^{1/2}}D\left(\epsilon _-\right)+\frac{\lambda
	^2}{\left|\epsilon _+\right|\left[\epsilon _+^2+\lambda ^2\right]{}^{1/2}}D\left(\epsilon _+\right)\text{d$\omega $}=2
\end{equation}

This implies that the sum of the oscillator strengths of the two lines at \(\epsilon _{\pm }=0\) is equal to 

\begin{equation}
I_{\epsilon }=2\left(1-I_{\mathcal{B}}\right)
\end{equation}

where

\begin{equation}
I_{\mathcal{B}}=\int _{-\infty }^{\infty }D(\mathcal{E})\sin ^2\theta  \text{d$\mathcal{E}$}
\end{equation}

is the oscillator strength of the line at \(\omega =\mathcal{B}\).

Assuming a Lorentian strain distribution (as suggested by Stoneham\cite{stoneham1975} in the low defect density limit)

\begin{equation}
D(\mathcal{E})=\frac{\gamma /\pi }{\mathcal{E}^2+\gamma ^2}
\end{equation}

where \(\gamma\) is the distribution width that is proportional to the defect density,

\begin{equation}
I_{\mathcal{B}}=\int _{-\infty }^{\infty }\frac{\gamma /\pi }{\mathcal{E}^2+\gamma ^2}\frac{\mathcal{E}^2}{\mathcal{E}^2+\lambda ^2}\text{d$\mathcal{E}$}=\frac{\gamma
}{\gamma +\lambda }\\
\\
\to I_{\epsilon }=\frac{2\lambda }{\gamma +\lambda }
\end{equation}

Thus, in the limit of a broad strain distribution \(\gamma >>\lambda\), the oscillator strengths are

\begin{equation}
I_{\mathcal{B}}\to 1, I_{\epsilon }\to 0
\end{equation}

Thus, for a typical ensemble sample, where the strain distribution is much broader than \(\lambda \sim\) GHz the EPR will only exhibit a single line at \(\omega =\mathcal{B}\). Resulting in a single that will be obscured by my prominent doublet spins in diamond, primarily the substitutional nitrogen or P1 center.